\documentclass{article}
\usepackage{todonotes}
\usepackage{PRIMEarxiv}
\usepackage{amsmath}
\usepackage[utf8]{inputenc} 
\usepackage{fancyvrb}
\usepackage{subcaption}

\usepackage{booktabs}
\usepackage{array}
\usepackage[T1]{fontenc}    
\usepackage{hyperref}       
\usepackage{url}    
\usepackage{tikz}
\usepackage{mdframed}
\usetikzlibrary{shapes.geometric, arrows} 
\usepackage{amsfonts}       
\usepackage{nicefrac}       
\usepackage{microtype}      
\usepackage{lipsum}
\usepackage{lscape}
\usepackage{longtable}
\usepackage{fancyhdr}       
\usepackage{colortbl}
\usepackage{makecell}
\usepackage{verbatim}
\usepackage{multirow}
\usepackage{ulem}
\usepackage{rotating}
\usepackage{caption}  

\usepackage{graphicx}       
\graphicspath{{media/}}     
\usepackage{ltcaption}
\usepackage{pdflscape}  
\usepackage{pgfplots}
\usepackage{filecontents}
\usepackage{pgfcalendar}
\usepackage{datetime}

\pgfplotsset{compat=1.17}
\usetikzlibrary{patterns, shapes.arrows, decorations.pathreplacing}
\usetikzlibrary{shapes.geometric, arrows, positioning}

\tikzstyle{startstop} = [rectangle, rounded corners, minimum width=3cm, minimum height=1cm,text centered, draw=red, fill=red!30]
\tikzstyle{process} = [rectangle, minimum width=3cm, minimum height=1cm, text centered, draw=blue, fill=blue!30]
\tikzstyle{arrow} = [thick,->,>=stealth]
\def\BibTeX{{\rm B\kern-.05em{\sc i\kern-.025em b}\kern-.08em
    T\kern-.1667em\lower.7ex\hbox{E}\kern-.125emX}}

\title{Leveraging Embedding Techniques in Multimodal Machine Learning for Mental Illness Assessment}

\author{ Abdelrahaman A. Hassan$^\dagger$,  Abdelrahman A. Ali$^\dagger$, Aya E. Fouda$^\dagger$, Radwa J. Hanafy$^{\ddagger,\dagger}$ and  Mohammed E. Fouda$^\dagger$ \\
$^\dagger$Compumacy for Artificial Intelligence solutions, Cairo, Egypt.\\
$^\ddagger$ Department of Behavioural Health- Saint Elizabeths Hospital, Washington DC, 20032.\\
fouda@compumacy.com
}

\begin{document}
\maketitle
\begin{abstract}
The increasing global prevalence of mental disorders, such as depression and PTSD, requires objective and scalable diagnostic tools.  Traditional clinical assessments often face limitations in accessibility, objectivity, and consistency.  This paper investigates the potential of multimodal machine learning to address these challenges, leveraging the complementary information available in text, audio, and video data. Our approach involves a comprehensive analysis of various data preprocessing techniques, including novel chunking and utterance-based formatting strategies.  We systematically evaluate a range of state-of-the-art embedding models for each modality and employ Convolutional Neural Networks (CNNs) and Bidirectional LSTM Networks (BiLSTMs) for feature extraction. We explore data-level, feature-level, and decision-level fusion techniques, including a novel integration of Large Language Model (LLM) predictions.  We also investigate the impact of replacing Multilayer Perceptron classifiers with Support Vector Machines.  We extend our analysis to severity prediction using PHQ-8 and PCL-C scores and multi-class classification (considering co-occurring conditions).  Our results demonstrate that utterance-based chunking significantly improves performance, particularly for text and audio modalities.  Decision-level fusion, incorporating LLM predictions, achieves the highest accuracy, with a balanced accuracy of 94.8\% for depression and 96.2\% for PTSD detection.  The combination of CNN-BiLSTM architectures with utterance-level chunking, coupled with the integration of external LLM, provides a powerful and nuanced approach to the detection and assessment of mental health conditions.  Our findings highlight the potential of MMML for developing more accurate, accessible, and personalized mental healthcare tools.
\end{abstract}


\section{Introduction}
\label{sec:introduction}

Mental disorders, including Post-Traumatic Stress Disorder (PTSD) and depression, represent a significant and growing global health challenge.  The World Health Organization (WHO) estimates that over one billion individuals worldwide suffer from mental illnesses, with a dramatic increase in anxiety and depression cases (over 25\%) observed during the first year of the COVID-19 pandemic \cite{who_covid_impact}.  PTSD, in particular, disproportionately affects individuals who have experienced traumatic events, posing substantial long-term risks to mental well-being and societal functioning \cite{PTSD_prevalence}.  The increasing prevalence of these disorders, coupled with limited access to mental healthcare professionals, creates an urgent need for scalable, objective, and accessible diagnostic tools.

Traditional clinical practices for psychiatric evaluation often suffer from subjectivity, inter-rater variability, and reliance on self-reported symptoms, which can be influenced by recall bias and stigma \cite{subjectivity_diagnosis}.  Furthermore, access to highly skilled mental health professionals is often limited by geographical constraints, cost, and long wait times \cite{access_mental_health}.  These limitations underscore the critical need for objective, reliable, and readily available methods for mental health assessment.

Artificial intelligence (AI), particularly machine learning (ML), has emerged as a transformative technology with the potential to address these challenges.  ML models can analyze complex, high-dimensional data from multiple sources (modalities) to identify subtle patterns indicative of mental health conditions, often surpassing human capabilities in detecting these nuanced signals \cite{hanafi2024comprehensive,8269806}. Recent advancements in multimedia technologies and ML have opened new avenues for automated medical diagnosis, offering the promise of timely and accurate identification of mental disorders, which could benefit millions worldwide.

Multimodal machine learning (MMML) offers a particularly promising approach.  By integrating information from different modalities, such as speech (audio), facial expressions (video), and written text, MMML can provide a more holistic and comprehensive assessment of an individual's mental state \cite{refNGUYEN20231458}.  For instance, studies have shown that visual cues like reduced emotional expressivity and altered eye movements can be strong indicators of depression.  Similarly, auditory features such as shortened speech segments, increased pause durations, and monotone delivery can be associated with depression and other mental health conditions .  By combining these diverse signals, MMML models can achieve a more robust and accurate diagnosis, capturing the complex interplay of symptoms that manifest across different channels.

This paper explores the effectiveness of MMML in detecting mental disorders, with a particular focus on depression and PTSD using the E-DAIC dataset \cite{Sadeghi2024, ringeval2019avec2019workshopchallenge}.  We introduce a novel, fully automated approach that leverages the power of Large Language Models (LLMs) and deep learning architectures to predict depression severity.  Our key contributions include:

\begin{enumerate}
    \item \textbf{Comprehensive Modality Analysis:} We conduct an in-depth investigation of various data preprocessing and formatting strategies for text, audio, and video modalities within the E-DAIC dataset, including novel chunking and utterance-based approaches.
    \item \textbf{Embedding Model Evaluation:}  We systematically evaluate a wide range of state-of-the-art embedding models, including both open-source and closed-source options, to identify the most effective representations for each modality.
    \item \textbf{Advanced Feature Extraction:} We employ a combination of Convolutional Neural Networks (CNNs) and Bidirectional Long Short-Term Memory networks (BiLSTMs) to extract rich, context-aware features from the embeddings, capturing both local and global patterns.
    \item \textbf{Fusion Strategies:} We explore and compare data-level, feature-level, and decision-level fusion techniques, including the integration of LLM-based predictions, to maximize the synergistic potential of multimodal data.  We also investigate the impact of replacing traditional Multilayer Perceptron (MLP) classifiers with Support Vector Machines (SVMs).
    \item \textbf{Severity and Multi-class Prediction:}  Beyond binary classification, we extend our analysis to include depression severity prediction (using PHQ-8 scores) and multi-class classification (considering co-occurring depression and PTSD), providing a more clinically relevant and nuanced assessment.
\end{enumerate}

Embeddings play a crucial role in our approach, transforming unstructured data (text, audio, video) into a structured, high-dimensional vector space suitable for machine learning models \cite{Bartal2023OpenAIsNE}.  In text analysis, word embeddings (e.g., Word2Vec, GloVe) capture semantic relationships between words \cite{article1}, while more recent LLMs generate contextual embeddings that represent entire sentences or documents, capturing nuanced meaning and context.  In the audio domain, embeddings represent acoustic features like tone, cadence, and speech patterns, which are known to be indicative of emotional states and mental health conditions \cite{ali2024leveragingaudiotextmodalities}. By converting diverse modalities into a unified representation, embeddings enable the integration of information from different sources, facilitating a more comprehensive and nuanced analysis of mental health indicators \cite{qin2025mentalperceiveraudiotextualmultimodallearning}.

The remainder of this paper is organized as follows: Section 2 reviews related work in multimodal mental health assessment. Section 3 details our methodology, including data preprocessing, embedding model selection, feature extraction, classification, and fusion techniques. Section 4 presents our experimental results and analysis. Section 5 discusses future directions for research.

\section{Related Work}
Mental health assessment using machine learning (ML) has gained significant traction, with researchers leveraging various modalities, such as text, audio, and video, to improve predictive performance. Different approaches range from single-modality studies to complex multimodal fusion techniques, aiming to extract meaningful patterns for mental illness classification. This section reviews the literature on mental health detection, focusing on dataset utilization, preprocessing strategies, feature extraction techniques, and classification models.

\subsection{E-DAIC Dataset and Its Modalities}
The E-DAIC dataset is one of the most widely used datasets in mental health detection. Its popularity stems from its multimodal nature, which provides researchers with text, audio, and video data to explore diverse analytical techniques. The dataset enables a more comprehensive study of depression and anxiety classification by allowing comparisons across different modalities or their combinations. Several studies have investigated single-modality approaches as well as multimodal fusion strategies to enhance classification accuracy.

\begin{itemize}
     \item \textbf{Text Modality:} A study by Sergio Burdisso et al.~\cite{Burdisso_2024} applied transformer-based language models to extract linguistic patterns indicative of depression and anxiety. The study focused on how textual cues within clinical interviews can be used to identify individuals suffering from depression, leveraging BERT-based embeddings to capture deep semantic meanings. Additionally, it investigated the impact of different linguistic features such as syntactic complexity and sentiment analysis in depression classification. The research highlighted that textual patterns, including sentence structure and word choice, can significantly contribute to predicting mental health conditions.
    
    \item \textbf{Audio Modality:} A study by Xiangsheng Huang et al.~\cite{Huang2024} proposed an AI-driven approach to depression recognition using voice-based pre-trained models. They employed the wav2vec 2.0 model as a feature extractor to automatically learn high-quality voice features from raw audio, eliminating the need for manual feature engineering. Their method was fine-tuned on the DAIC-WOZ dataset, achieving state-of-the-art classification performance. The study highlighted that voice-based deep learning models can significantly improve depression detection by leveraging pre-trained representations of speech, thereby enhancing the generalization ability of depression classifiers. 

    \item \textbf{Video Modality:} A study by Avantika Shrestha et al.~\cite{10386191} explored the use of temporal facial features extracted from clinical interviews to screen for major depressive disorder (MDD) and PTSD. Their multi-task learning framework utilized a bidirectional GRU with self-attention, demonstrating that incorporating temporal facial features improved generalization performance over single-task approaches.
    \item \textbf{Multimodal Fusion:} Chayan Tank et al.~\cite{tank2024depressiondetectionanalysisusing} proposed a framework combining text and audio features using deep learning-based fusion strategies, demonstrating improved accuracy over single-modality baselines. Another study by Santosh V. Patapati et al.~\cite{patapati2024integratinglargelanguagemodels} explored a more comprehensive multimodal model incorporating additional behavioral cues alongside text and audio, leveraging cross-modal attention mechanisms to enhance feature interactions.
\end{itemize}

\subsection{Preprocessing and Feature Extraction}
Different preprocessing techniques and models have been used to handle various modalities in mental health assessment. Some studies rely on traditional methods such as normalization, noise removal, and handcrafted feature extraction, while others utilize pre-trained deep learning models to learn representations directly from raw data. The effectiveness of preprocessing significantly impacts the quality of extracted features and, consequently, the accuracy of classification models. Here are some examples of preprocessing techniques applied to different modalities:

\subsubsection{Text Modality}
\begin{itemize}
    \item \textbf{Standard Text Normalization}
     Lorenzoni et al.~\cite{lorenzoni2024assessingmlclassificationalgorithms} applied fundamental text preprocessing techniques such as punctuation removal, conversion to lowercase, and stopword elimination. These steps aimed to reduce noise and ensure consistency across samples for more effective feature extraction. Additionally, their study filtered artificial conversation markers present in the dataset, improving the authenticity of the textual data used for classification.
     \item \textbf{TF-IDF Feature Extraction}  
    Mariia Nykoniuk et al.~\cite{computation13010009} utilized Term Frequency-Inverse Document Frequency (TF-IDF) as a feature extraction technique to quantify the importance of words within the dataset. This method assigns higher weights to words that appear frequently in a given document but are rare across the entire corpus, allowing models to focus on depression-related keywords. Their approach demonstrated that TF-IDF can enhance text-based depression detection by capturing significant linguistic patterns.

     \item \textbf{Embedding-Based Feature Extraction}
     Lau et al.~\cite{Lau2023} leveraged transformer-based models for more advanced text encoding. Their approach included preserving punctuation and transcribing non-verbal sounds such as laughter and sighs, recognizing their importance in identifying depressive speech patterns. Additionally, they concatenated interviewer questions with participant responses to maintain conversational context, ensuring that critical dependencies were preserved. Furthermore, they optimized text encoding through parameter-efficient fine-tuning, demonstrating improvements over traditional fine-tuning methods and enhancing classification accuracy.
\end{itemize}

\subsubsection{Audio Modality}
\begin{itemize}
    \item \textbf{Traditional Feature Engineering}
    Rohan Gupta et al.~\cite{GUPTA2025101710} applied various preprocessing techniques to enhance speech-based feature extraction. These included noise reduction, voice activity detection, and mel-spectrogram transformations. Additionally, they extracted prosodic and spectral features to differentiate depressive and non-depressive speech patterns, focusing on changes in pitch, intensity, and rhythm as potential indicators of depression.
    \item \textbf{Embedding-Based Feature Extraction}
    Huang et al.~\cite{Huang2024} leveraged the wav2vec 2.0 model, a voice-based pre-trained deep learning model, to extract high-quality speech features directly from raw audio. This approach eliminated the need for manual feature engineering, allowing the model to learn representative embeddings in a self-supervised manner. Their method demonstrated strong generalization ability, achieving high classification accuracy in both binary and multi-class depression detection tasks.
\end{itemize}

\subsubsection{Video Modality}
\begin{itemize}
    \item \textbf{Traditional Feature Engineering}
    A study by Baydili et al.~\cite{bs15030352} applied deep learning techniques for depression and suicidal tendency detection from video-based social media data. The research utilized feature selection techniques and pre-trained neural networks to extract meaningful patterns from facial expressions and movement dynamics. Their method demonstrated the effectiveness of automated facial feature extraction in detecting mental health conditions.
    \item \textbf{Deep Learning-Based Feature Extraction}
    Chayan Tank et al.~\cite{tank2024depressiondetectionanalysisusing} employed advanced deep learning techniques for face detection, landmark extraction, and feature normalization. Their approach included the extraction of facial action units (FAUs) and microexpressions while normalizing variations in lighting and pose to ensure consistent visual feature representation. Their method enhanced the robustness of facial expression analysis for mental health assessment, contributing to improved model generalization.
\end{itemize}

\subsection{Classifiers}
\label{classifiers}

Various types of models have been employed to analyze mental health data, ranging from traditional machine learning techniques to deep learning and LLMs. Giuliano Lorenzoni et al.~\cite{lorenzoni2024assessingmlclassificationalgorithms} utilized classical ML techniques such as SVMs and random forests for mental health classification tasks. These models rely on handcrafted feature extraction and have demonstrated competitive performance in specific scenarios, particularly when training data is limited. Deep learning has played a significant role in advancing mental health assessments. In a study by Rohan Gupta et al.~\cite{GUPTA2025101710}, deep learning models were employed to process multimodal data, capturing complex patterns from speech, text, and video features. Notably, MLPs were used as the primary classification model, effectively learning non-linear feature representations for depression detection. The study demonstrated that MLPs, in combination with deep feature extraction, significantly improved classification accuracy compared to traditional ML techniques. More recently, LLMs have been employed to analyze textual and multimodal mental health data. A study by Bakir Hadzic et al.~\cite{Hadzic31122024} demonstrated the effectiveness of LLMs like GPT in understanding contextual cues from patient dialogues. By leveraging transfer learning, these models have achieved state-of-the-art results in depression and anxiety detection, highlighting the growing role of LLMs in this domain.

\subsection{Fusion Techniques}

Multimodal fusion techniques have been widely explored on the E-DAIC dataset to enhance predictive performance in mental health assessment. Several approaches integrate different modalities at various levels to maximize information extraction and classification accuracy.

One approach, as presented by Mariia Nykoniuk et al.~\cite{computation13010009}, employs intermediate fusion, where features from different modalities are combined at a feature representation level before classification. Their method integrates text and audio embeddings using attention-based fusion techniques to capture meaningful cross-modal dependencies, resulting in improved depression classification.

Another study by Santosh V. Patapati et al.~\cite{patapati2024integratinglargelanguagemodels} explores cross-modal fusion using transformer-based architectures. By leveraging self-attention mechanisms, their model dynamically weights features from text, audio, and video, allowing for more flexible and robust mental health predictions. The integration of multiple modalities through cross-attention modules significantly improves generalization performance across different patient populations.

\subsection{Evaluation Metrics}

In the assessment of machine learning models for mental health prediction, different evaluation metrics are employed to capture various aspects of model performance. Since mental health assessment tasks involve different types of predictions, including binary classification, severity prediction, and multiclass classification, different metrics are required to measure performance effectively. Below, we discuss metrics we used in these tasks.

\textbf{Balanced Accuracy (BA)}
With imbalanced datasets, balanced accuracy accounts for the model’s ability to correctly classify both positive and negative cases by averaging the recall values of each class. The balanced accuracy for binary classification is defined as:

\begin{equation}
\text{\textbf{Balanced Accuracy}} = \frac{1}{2} \left( \frac{TP}{TP + FN} + \frac{TN}{TN + FP} \right)
\end{equation}
where TP, TN, FP and FN stand for True Positives, True Negatives, False Positives and False Negatives, respectively.

For \textbf{multiclass classification}, balanced accuracy generalizes as:

\begin{equation}
\text{\textbf{Balanced Accuracy}} = \frac{1}{N} \sum_{i=1}^{N} \frac{\text{TP}_i}{\text{TP}_i + \text{FN}_i}
\end{equation}
where $N$ is the number of classes, and $TP_i$ and  $FN_i$ correspond to the true positive and false negative counts for each class $i$, respectively.

Balanced accuracy is particularly beneficial in mental health classification tasks, where datasets often suffer from class imbalances, making standard accuracy less informative.

\textbf{Mean Absolute Error (MAE)} is a commonly used metric for \textbf{regression tasks}, such as predicting the severity score of mental health disorders. It calculates the absolute differences between predicted and actual values, making it easy to interpret. The formula for MAE is:

\begin{equation}
MAE = \frac{1}{n} \sum_{i=1}^{n} |y_i - \hat{y}_i|
\end{equation}
where $y_i$ is the actual severity score. $\hat{y}_i$ is the predicted severity score. And, $n$ is the number of samples. A lower MAE value indicates more precise predictions. In mental health assessments, MAE is useful for estimating the severity of conditions like depression or anxiety based on multimodal inputs. This metric is particularly relevant in predicting mental health severity scores, where slight deviations can significantly impact clinical decisions.

These evaluation metrics provide a comprehensive framework for assessing the performance of machine learning models in mental health applications. While balanced accuracy is particularly useful for classification problems (both binary and multiclass), MAE offers insights into the quality of regression predictions, making them valuable in tasks involving severity estimation. Future research should focus on refining these metrics to align more closely with real-world clinical decision-making in mental health diagnosis and treatment.

\section{Dataset}
\label{sec:dataset}
The Enhanced Distress Analysis Interview Corpus (E-DAIC) \cite{ringeval2019avec2019workshopchallenge} expands upon the original DAIC-WOZ dataset, focusing on psychological disorders such as depression and PTSD through structured semi-clinical interviews. These interviews take place within a wizard-of-Oz (WoZ) framework, where a human-operated virtual agent named "Ellie" conducts the sessions. The dataset includes text, audio, and video modalities, enabling comprehensive analysis of verbal and non-verbal cues and serving as a robust foundation for developing and evaluating multimodal models for mental health diagnostics.

A multimodal approach is utilized to process the dataset, incorporating text, audio, and video data. For text, the Whisper v3 model \cite{radford2022whisper} transcribes and segments the interviews into utterances, producing both the transcription and corresponding timestamps (start and end times). However, the initial transcriptions contain inaccuracies and repetitions. To refine the text data, Gemini 2.0 \cite{googleai2024studio} is used to detect and correct transcription errors and remove duplicated utterances. The cleaned transcriptions, along with their adjusted timestamps, then serve as a basis for extracting corresponding audio segments. For the audio modality, both the raw full interview recordings and the extracted audio segments, based on the cleaned timestamps, undergo further processing. Meanwhile, for the video modality, the pre-extracted OpenFace features provided in the dataset are utilized.

The E-DAIC dataset consists of 275 interview samples, divided into training, development, and test sets.

\subsection{Class Distributions}

The E-DAIC dataset encompasses a range of clinical presentations, categorized as follows:

\textbf{Depression Binary:} The E-DAIC dataset includes a binary classification for depression, where participants are categorized as either depressed or non-depressed. A participant is labeled as depressed if their PHQ-8 score is higher or equal to 10. The dataset contains 189 depressed and 86 non-depressed samples. Notably, 20 samples were originally mislabeled as "Negative" for depression despite having PHQ-8 scores exceeding 10. These mislabeled samples correspond to IDs: [320, 325, 335, 344, 352, 356, 380, 386, 409, 413, 418, 422, 433, 459, 483, 633, 682, 691, 696, 709].

\textbf{Depression Severity:} In addition to the binary classification, the dataset also provides a severity assessment for depression. This categorization is based on PHQ-8 scores, following the methodology outlined by Kroenke et al. \cite{articles}. The severity levels are defined as follows:
\begin{itemize}
    \item Minimal: 0-4 (122 samples)
    \item Mild: 5-9 (67 samples)
    \item Moderate: 10-14 (43 samples)
    \item Moderately Severe: 15-19 (33 samples)
    \item Severe: 20-24 (10 samples)
\end{itemize}

\textbf{PTSD Binary:} The dataset also includes a binary classification for PTSD. Participants are classified as either PTSD-positive or PTSD-negative. A participant is labeled as PTSD-positive if their PCL-C score is higher than 44. The dataset comprises 188 PTSD-positive and 87 PTSD-negative samples.

\textbf{PTSD Severity:} Furthermore, the severity of PTSD is categorized based on PCL-C scores, following the guidelines from García-Valdez et al. \cite{10.1007/978-3-031-46933-6_21}. The severity levels are as follows:
\begin{itemize}
    \item Little or no severity: 17-29 (137 samples)
    \item Moderate severity: 30-44 (51 samples)
    \item High severity: 45-85 (87 samples)
\end{itemize}

\subsection{Data Reprocessing}

\subsubsection{Whole Interview Format}

In the whole interview format, data across all modalities is structured so that each interview is represented by a single embedding vector. For the text modality, two approaches are tested. The first approach uses the entire transcript of the interview as input to the embedding model, while the second approach restructures the text into a question-and-answer format, as shown in Fig \ref{qa_format}, which is extracted using regular expressions.

\begin{figure}[ht]
    \centering
    \begin{mdframed}[linewidth=1pt] 
        \begin{minipage}{0.8\textwidth}
            \begin{Verbatim}
Question 1: How are you doing today?
Answer 1: Good.

Question 2: What's good? Where are you from originally?
Answer 2: Atlanta, Georgia.

Question 3: Really? Why'd you move to LA?
Answer 3: My parents are from here.

Question 4: How do you like LA?
Answer 4: I love it.

Question 5: What are some things you really like about LA?
Answer 5: I like the weather. I like the opportunities.
            \end{Verbatim}
        \end{minipage}
    \end{mdframed}
    \caption{Interview data example as QA Format.}
    \label{qa_format}
\end{figure}

For the audio modality, two processing strategies are adopted. The first approach retains the full raw audio of the interview to capture vocal nuances and inflections throughout the session. The second approach relies on the cleaned timestamps from Whisper, extracting only the participant's responses to exclude the interviewer's speech. This method focuses on isolating the verbal cues directly relevant to the participant's mental state.

\subsubsection{Chunking Format}
To refine data preprocessing further, a chunking strategy is employed, where groups of consecutive utterances are treated as discrete input chunks. Each chunk is considered an individual timestep, allowing the model to process sequential interactions independently. To preserve conversational context, an overlapping sliding window approach is applied, ensuring that follow-up questions remain connected to their preceding context. This method captures the conversational flow and emotional or thematic shifts within the interviews, maintaining coherence in the dialogue.

For the text modality, the interview is first segmented into utterances using Whisper v3, which provides a transcript along with the start and end timestamps for each utterance. Instead of directly chunking the textual data, the timestamps are grouped into chunks using multiple techniques. These chunked timestamps are then used to extract the corresponding text segments, ensuring that each chunk preserves the natural structure of the conversation while maintaining alignment with other modalities.

For the audio modality, the same chunking strategy is adopted, using the previously grouped chunked timestamps to extract the corresponding utterance-level audio segments. This ensures that the entire conversation, including both the interviewer’s questions and the participant’s responses, is segmented in a manner that aligns with the text chunks. By mirroring the text chunking approach, the interplay between verbal content, tone, and intonation is preserved, enhancing the model's ability to recognize speech patterns and emotional expressions.

For the video modality, since the dataset does not include raw videos but instead provides pre-extracted OpenFace features for each frame, these frame-level features are processed using an extraction and averaging method. Specifically, the maximum value of all extracted features within each utterance is computed to generate a single representative feature per utterance. These max-pooled features are then structured into timestep-aligned chunks based on the chunked timestamps, ensuring that each video feature chunk corresponds to its textual and audio counterparts. This structured approach preserves essential facial expression and visual cue information within the conversation.
This multimodal preprocessing pipeline ensures that textual, auditory, and visual data are consistently formatted and aligned, thereby improving the model’s ability to analyze both verbal and non-verbal cues in mental health assessments.

\section{Methodology}

This section outlines the methodology employed for analyzing the E-DAIC dataset and developing a multimodal framework for mental health diagnostics. As shown in Figure \ref{fig:workflow}, the process begins with data preprocessing, which includes transcription refinement and segmentation. Two primary data formatting strategies are adopted: the Whole Interview Format, which retains the entire session as a single sequence, and the Chunking Format, which segments the interview into smaller units for finer-grained analysis. We then evaluate multiple embedding models across the text, audio, and video modalities to identify the most effective feature representations. The subsequent stages involve feature extraction, model training, and multimodal fusion, followed by downstream tasks such as severity prediction and multi-class classification. This end-to-end pipeline is designed to fully leverage multimodal data for robust mental health assessment.

\begin{figure}
    \centering
    \includegraphics[width=1\linewidth]{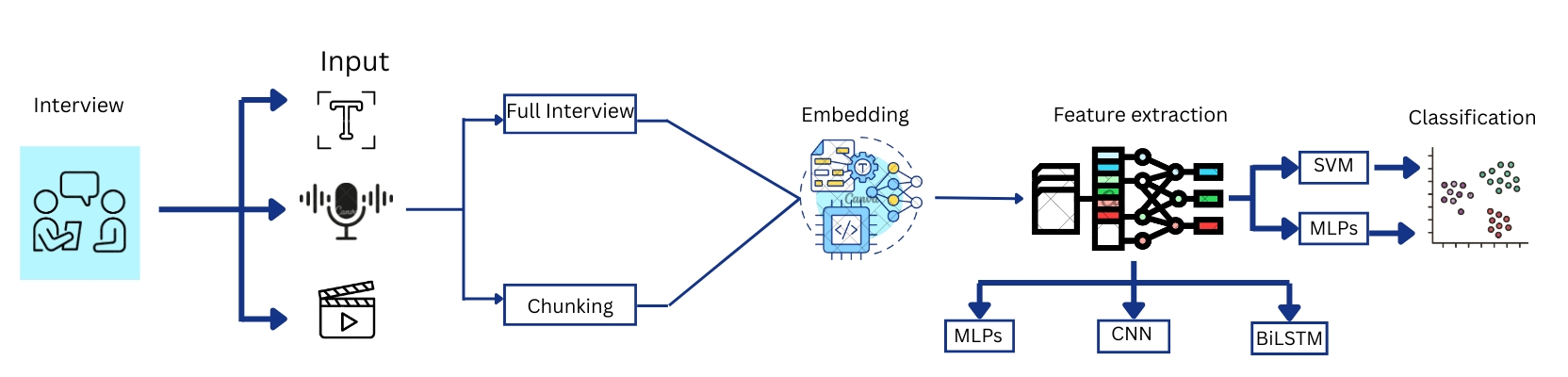}
    \caption{Proposed Pipeline for Interview Analysis}
    \label{fig:pipeline}
\end{figure}

Figure \ref{fig:pipeline} illustrates the proposed processing pipeline, which is applied independently to each modality (i.e., text, audio, and video). As described earlier, the input interview data is first preprocessed and formatted based on the selected strategy (full interview or chunked format). The formatted data is then converted into embeddings using modality-specific models, producing fixed-size vector representations. These embeddings are passed into feature extraction architectures, such as Multi-Layer Perceptrons (MLPs), CNNs, or Bidirectional Long Short-Term Memory networks (BiLSTMs), each tailored to capture salient patterns within the data. Finally, the extracted features are fed into classification layers—typically MLPs or SVMs—to perform predictive tasks relevant to mental health evaluation.

\begin{figure}[ht]
\large
    \centering
\begin{mdframed}[backgroundcolor=gray!10, linecolor=gray!10]
\begin{center}
\resizebox{0.9\textwidth}{!}{%
\begin{tikzpicture}[node distance=2cm]

\tikzstyle{process} = [rectangle, rounded corners, minimum width=3cm, minimum height=1cm, text centered, draw=black, fill=blue!20]
\tikzstyle{startstop} = [rectangle, rounded corners, minimum width=3cm, minimum height=1cm, text centered, draw=black, fill=teal!80, text=white]
\tikzstyle{arrow} = [very thick,->,>=stealth]  

\node (data) [startstop] {Data};
\node (embedding) [startstop, right of=data, xshift=3cm] {Embedding};
\node (feature) [startstop, right of=embedding, xshift=3cm] {Feature Extraction};
\node (classification) [startstop, right of=feature, xshift=3cm] {Classification};
\node (fusion) [startstop, right of=classification, xshift=3cm] {Fusion};

\node (full) [process, below of=data, yshift=-0.5cm, xshift=1cm] {Full Interview};
\node (chunk) [process, below of=full, yshift=-0.5cm] {Chunking};

\node (text) [process, below of=embedding, yshift=-0.5cm, xshift=1cm] {Text Models};
\node (audio) [process, below of=text, yshift=-0.5cm] {Audio Models}; 
\node (video) [process, below of=audio, yshift=-0.5cm] {Video Models};

\node (mlp) [process, below of=feature, yshift=-0.5cm, xshift=1cm] {MLP};
\node (cnn) [process, below of=mlp, yshift=-0.5cm] {CNN};
\node (bilstm) [process, below of=cnn, yshift=-0.5cm] {BiLSTM};

\node (mlp_class) [process, below of=classification, yshift=-0.5cm, xshift=1cm] {MLP};
\node (svm) [process, below of=mlp_class, yshift=-0.5cm] {SVM};

\node (data_level) [process, below of=fusion, yshift=-0.5cm, xshift=1cm] {Data-Level};
\node (feature_level) [process, below of=data_level, yshift=-0.5cm] {Feature-Level};
\node (decision_level) [process, below of=feature_level, yshift=-0.5cm] {Decision-Level};

\draw [arrow] (data.south) ++(-1cm,0) |- (full.west);
\draw [arrow] (data.south) ++(-1cm,0) |- (chunk.west);

\draw [arrow] (embedding.south) ++(-1cm,0) |- (text.west);
\draw [arrow] (embedding.south) ++(-1cm,0) |- (audio.west); 
\draw [arrow] (embedding.south) ++(-1cm,0) |- (video.west);

\draw [arrow] (feature.south) ++(-1cm,0) |- (mlp.west);
\draw [arrow] (feature.south) ++(-1cm,0) |- (cnn.west);
\draw [arrow] (feature.south) ++(-1cm,0) |- (bilstm.west);

\draw [arrow] (classification.south) ++(-1cm,0) |- (mlp_class.west);
\draw [arrow] (classification.south) ++(-1cm,0) |- (svm.west);

\draw [arrow] (fusion.south) ++(-1cm,0) |- (data_level.west);
\draw [arrow] (fusion.south) ++(-1cm,0) |- (feature_level.west);
\draw [arrow] (fusion.south) ++(-1cm,0) |- (decision_level.west);

\draw [arrow] (data) -- (embedding);
\draw [arrow] (embedding) -- (feature);
\draw [arrow] (feature) -- (classification);
\draw [arrow] (classification) -- (fusion);

\end{tikzpicture}
}
        \caption{Process Workflow}
    \label{fig:workflow}
\end{center}
\end{mdframed}
\end{figure}

\subsection{Embedding Models Evaluation} \label{emb_eval}
This section details the evaluation of embedding models used to represent the three modalities of the data: audio, text, and video. A critical step in the multimodal analysis pipeline involves identifying and configuring the most effective embedding models for each modality to ensure accurate and informative representations. The evaluation focuses on the ability of each embedding type to capture salient features relevant to the classification task, with performance assessed using SVM classifiers and optimized through hyperparameter tuning with Optuna. The following subsections describe the specific models and evaluation processes applied to each modality.

\subsubsection{Audio Embeddings}
Audio data is transformed into embeddings using pre-trained feature extractors, including Whisper and wav2vec. These models are evaluated to determine the most suitable one for capturing both temporal and frequency-based features. The embeddings are extracted from the last layer of the encoder, ensuring that they contain rich contextual information. The selected model, which provides the most comprehensive representations, is integrated into the processing pipeline. The performance of these audio embeddings is assessed using SVM\cite{cortes1995support}, with additional optimization performed through high-parameter search techniques such as Optuna to achieve the best classification metrics.
\subsubsection{Text Embeddings}
For textual data, transformer-based embedding models are tested to identify the most effective representation method. The text data originates from transcriptions of the interviews, preprocessed for consistency. Whisper, a state-of-the-art automatic speech recognition (ASR) model, is used to transcribe the audio data, ensuring high transcription quality. These transcriptions are then converted into embeddings using the best-performing text embedding model, selected through a comprehensive evaluation. The resulting text embeddings are further analyzed using SVM classifiers, with hyperparameter tuning via Optuna to optimize classification performance.

\subsubsection{Video Embeddings}

Video features are extracted directly from the dataset, emphasizing motion dynamics and visual patterns. OpenFace is utilized to capture key facial expressions, gestures, and other relevant visual cues. Key frames are selected from the video data, and the extracted features are encoded into embeddings. The evaluation of video embeddings focuses on their effectiveness in modeling motion and interaction cues within the dataset. These embeddings are integrated with text and audio embeddings using different fusion strategies—data-level, feature-level, and decision-level fusion—and analyzed for their impact on classification performance. The results of this evaluation, presented in the Results section, highlight the contribution of motion-specific features to overall system performance.

\subsection{Feature Extraction}
\label{feat_ex}
After the embedding model evaluation, we selected the best-performing models from the experiment to serve as the foundation for feature extraction. Feature extraction is a critical step in multimodal analysis, providing structured insights from diverse inputs such as text, audio, and video. Unlike approaches that rely solely on pre-trained models, we trained custom neural network models directly on the E-DAIC dataset to capture patterns and nuances specific to this domain. This approach ensured that the feature extraction process was fine-tuned to the unique characteristics of the dataset, enabling the detection of subtle emotional or behavioral patterns. By tailoring neural networks to the dataset, we enhanced performance across multiple modalities, uncovering intricate patterns that pre-trained models might overlook. This pipeline emphasized the importance of domain-specific training, allowing the models to leverage the embeddings effectively for downstream tasks.

For the whole interview format, where each interview was represented by a single embedding vector, MLPs were employed to model non-linear relationships and capture high-level representations across text, audio, and video. This approach allowed the model to leverage the full context of each interview holistically, ensuring a comprehensive understanding of the interactions within the session.

In the chunking format, where data was structured as a two-dimensional representation with time steps (chunks) along one dimension and the embedding vector along the other, CNNs were used to extract spatial and temporal patterns. For audio embeddings, CNNs captured intricate frequency and intensity relationships, while for video embeddings, CNNs identified meaningful patterns related to facial expressions and micro-movements indicative of emotional states.

To further enhance the temporal modeling capabilities in the chunking format, CNN-BiLSTM hybrid architectures were implemented. By combining the spatial feature extraction strengths of CNNs with the sequential modeling power of BiLSTMs, these models effectively captured complex temporal-spatial relationships, particularly in the audio and video modalities. This comprehensive feature extraction framework ensured that rich and meaningful features were derived from the embeddings, optimizing the multimodal analysis for improved robustness and classification accuracy.

\subsection{Classification}
\label{class}
In this study, we explored the potential of SVMs to improve the accuracy of MLP classifiers. To test this, we incorporated an SVM into our workflow by using deep learning models such as CNNs or BiLSTMs for feature extraction. Specifically, during training, we used the MLP layer as part of the deep learning architecture. For inference, the model was modified to output the features produced either by the convolutional layer of the CNN or the last hidden layer of the BiLSTM. These features were extracted for both the training and test datasets.

Subsequently, an SVM was trained on the features extracted from the training set and tested on the features from the test set. This hybrid approach allowed us to leverage the strengths of both models: deep learning models excel at capturing complex spatial and temporal features, while SVMs are robust classifiers that excel at handling high-dimensional data and finding optimal decision boundaries, even in cases where the data is not linearly separable. By integrating the SVM, we aimed to improve classification accuracy and generalization, particularly in scenarios with limited or imbalanced datasets. This approach highlighted the potential of SVMs to enhance the performance of traditional MLP classifiers in the context of mental health diagnostics. 

\subsection{Data Fusion}

\label{fusion}
Data fusion is a critical step in multimodal analysis, as it involves integrating information from different modalities (text, audio, and video) to enhance the overall performance and robustness of the model. In this study, we tested three primary fusion strategies: data-level fusion, feature-level fusion, and decision-level fusion. Each strategy operates at a different stage of the pipeline and offers unique advantages depending on the nature of the data and the specific task.

\subsubsection{Data-Level Fusion}
\begin{figure}
    \centering
    \begin{subfigure}[b]{1\textwidth}
        \centering
        \includegraphics[width=\textwidth]{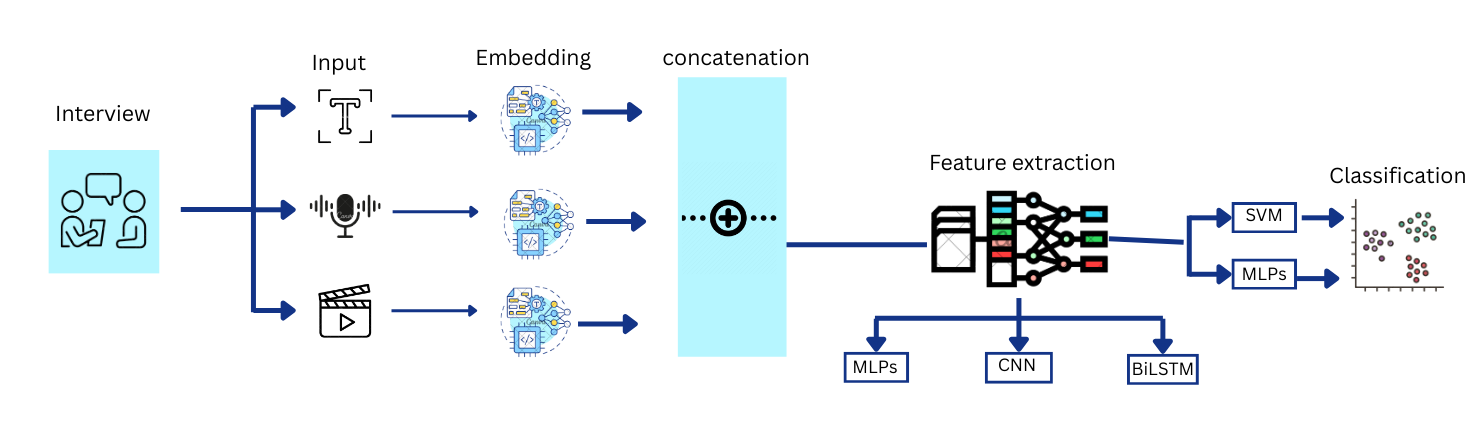}
        \caption{Data-Level Fusion}
        \label{fig:Data-Level Fusion.png}
    \end{subfigure}
    \hfill
    \begin{subfigure}[b]{1\textwidth}
        \centering
        \includegraphics[width=\textwidth]{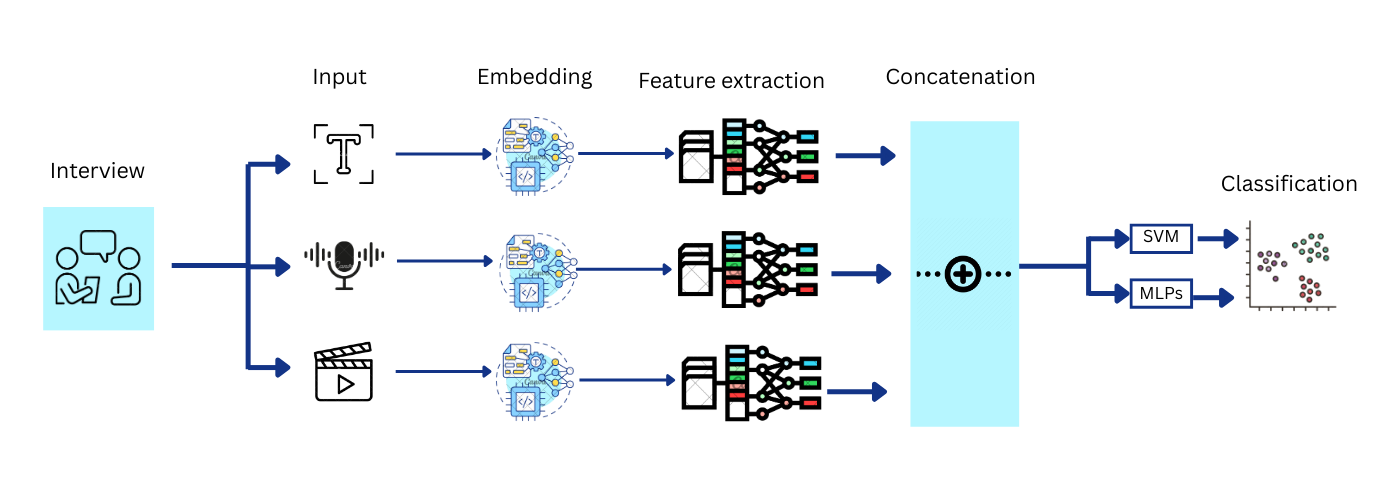}
        \caption{Feature-Level Fusion}
        \label{fig:Feature_Level_Fusion}
    \end{subfigure}
    \hfill
    \begin{subfigure}[b]{1\textwidth}
        \centering
        \includegraphics[width=\textwidth]{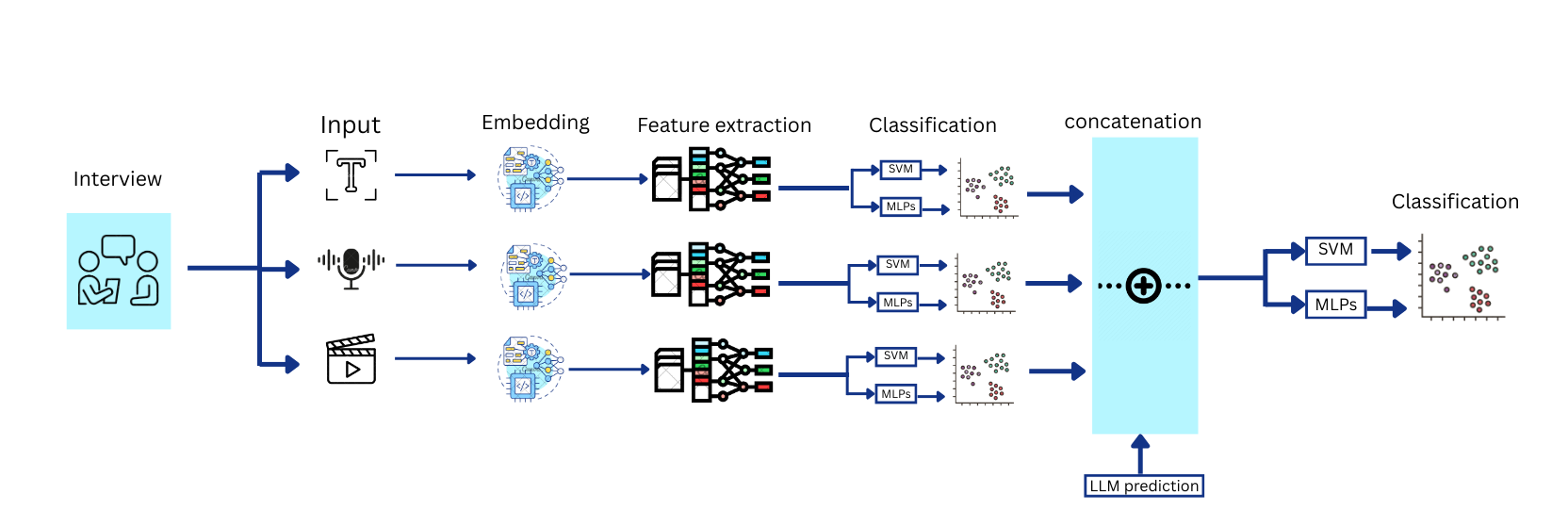}
        \caption{Decision-Level Fusion}
        \label{fig:Decision_Level_Fusion}
    \end{subfigure}
    
    \caption{Fusion strategies in multi-modal data processing: (a) Data-Level Fusion - Multi-modal interview data (text, audio, video) undergoes embedding, fusion, and advanced feature extraction using MLPs, CNN, and BiLSTM before classification with SVM and MLPs. (b) Feature-Level Fusion - Extracted features from different modalities are combined for learning. (c) Decision-Level Fusion - Separate model outputs are integrated for final prediction.}
    \label{fig:Fusion_Strategies}
\end{figure}

In Data-Level Fusion, preprocessed data from different modalities were combined before feature extraction. Unlike traditional approaches that require padding, we concatenated the features along the embedding dimension, ensuring that all modalities remained aligned based on the Whisper-generated timestamps. This approach was feasible because the data had synchronized timesteps across text, audio, and video, allowing for a direct concatenation of features without altering their temporal structure.

The fused representation was then passed through the best-performing model pipeline, which included steps such as normalization (when applicable) and classification using either MLPs or SVMs. By applying data-level fusion at this stage, the model was able to learn low-level interactions between modalities, capturing synergistic patterns that might otherwise be lost in later processing. This method leveraged the complementary nature of different data types while preserving the structural integrity of multimodal inputs. Figure \ref{fig:Data-Level Fusion.png} shows the process that we followed to develop Data-Level Fusion .

\subsubsection{Feature-Level Fusion}

In Feature-Level Fusion ,involved combining the high-dimensional feature embeddings extracted from each modality. For this, we utilized the best-performing model pipeline for text, audio, and video to generate modality-specific feature outputs. These outputs were then concatenated to create a single representation of the multimodal features. This combined feature set was used to train either a MLP or a SVM for classification. Additionally, we tested an end-to-end approach where the model architecture included separate layers for each modality, and the concatenation occurred within the model. This allowed the network to jointly learn modality-specific features while optimizing the fusion process in the same training cycle. This strategy was particularly effective for capturing complementary information across modalities, enabling the model to identify intricate patterns that were not evident in isolated embeddings.Figure \ref{fig:Feature_Level_Fusion} shows the process that we followed to develop Feature-Level Fusion

\subsubsection{Decision-Level Fusion}


In Decision-Level Fusion aggregated the outputs of modality-specific models at the final prediction stage, ensuring that each modality contributed independently to the classification decision. Separate models were trained for each modality, and their outputs were fused to generate a comprehensive final prediction. Additionally, we incorporated LLMs to generate text-based predictions, selecting the best-performing LLM for each task as identified in \cite{ali2024leveragingaudiotextmodalities}. This ensured that the most effective LLM was utilized for extracting meaningful insights from interview transcripts. Figure \ref{fig:Decision_Level_Fusion} shows the process that we followed to develop Decision Level Fusion

Two main approaches were employed for decision fusion:
\begin{itemize}
    \item \textbf{Logit Fusion:} The probability outputs (logits) from the modality-specific models were concatenated and passed through an SVM classifier, which determined the final prediction based on the combined confidence scores.
    
    \item \textbf{Binary Prediction Fusion:} The binary predictions (class labels) from each model, including text, audio, and video classifiers, as well as the LLM-based prediction, were concatenated and input into an SVM for final classification. By incorporating LLMs as part of the text-based prediction, we leveraged the model's ability to analyze contextual and linguistic cues, enhancing the robustness of the fusion process.
\end{itemize}

These fusion strategies allow independent predictions from each modality while optimizing classification accuracy through aggregated decision-making. Decision-level fusion was particularly beneficial in cases where modalities were less correlated, ensuring that each input contributed uniquely to the overall system performance. The integration of LLM-based predictions further strengthened the model’s ability to detect nuanced language patterns indicative of mental health conditions, adding an additional layer of interpretability and reliability to the classification framework.

\subsection{Severity and Multi-class Classification Tasks}
\label{sev}
Beyond binary classification for depression and PTSD, severity prediction and multi-class classification were evaluated at both the feature extraction and fusion levels. This broader analysis allowed for assessing the model’s capability to capture nuanced distinctions in mental health conditions beyond the binary presence or absence of a disorder.

For severity prediction, depression and PTSD severity scores provided in the dataset were used as regression targets. Models were evaluated not only using features extracted at the feature extraction stage but also at the fusion level, where different modalities were combined. Performance was measured using Mean Absolute Error (MAE) to quantify the deviation between predicted and actual severity scores, providing insight into the model’s ability to estimate mental health severity on a continuous scale.

For multi-class classification, the binary classification framework was expanded to distinguish among four distinct mental health conditions:
\begin{enumerate}
\item Neither depression nor PTSD
\item Depression only
\item PTSD only
\item Both depression and PTSD
\end{enumerate}

Multi-class classification was performed using models trained at both the feature extraction and fusion levels, ensuring a comprehensive evaluation of their effectiveness. Performance was assessed using accuracy, macro-F1 score, precision, recall, and confusion matrices, offering a detailed understanding of the model’s ability to distinguish between different mental health conditions.

By evaluating severity prediction and multi-class classification at multiple levels, the analysis provided a more thorough validation of the proposed approach. This assessment reinforced the robustness of the models in capturing complex mental health states, demonstrating their applicability in both categorical diagnosis and continuous severity estimation.

\section{Results}

\subsection{Data Formatting}
This section examines the impact of different data formatting strategies on classification performance for depression and PTSD across text, audio, and video modalities. This evaluation aims to identify the most effective formatting techniques for each modality, which are then used in subsequent classification tasks. Various chunking methods, feature selection approaches, and normalization techniques are tested to assess their influence on the accuracy of the model. The following subsections detail the results for each modality, highlighting the most effective formatting strategies.
\begin{table}[ht]
\centering

\caption{ Balanced accuracy evaluation of data formats across modalities on the test set. Values in bold represent the highest balanced accuracy achieved for each modality.}

\label{tab:format-eval}
\begin{tabular}{|ll|c|cc|cc|}
\hline
\multicolumn{2}{|c|}{\multirow{2}{*}{Approach}} &
  \multicolumn{1}{l|}{\multirow{2}{*}{Model}} &
  \multicolumn{2}{c|}{Depression} &
  \multicolumn{2}{c|}{PTSD} \\ \cline{4-7} 
\multicolumn{2}{|c|}{}                                                  & \multicolumn{1}{l|}{}   & BA   & BA (w/ norm) & BA   & BA (w/ norm) \\ \hline
\multicolumn{1}{|l|}{\multirow{7}{*}{\begin{turn}{90}Text\end{turn}}}  & Full Interview           & \multirow{2}{*}{SVM}    & 76.2 & 71.9    & 78.6 & 70      \\
\multicolumn{1}{|l|}{}                       & QA                       &                         & 77.6 & 78.5    & 80   & 74.8    \\ \cline{2-7} 
\multicolumn{1}{|l|}{}                       & 5 QA pairs               & \multirow{5}{*}{BiLSTM} & 81.9 & 86.2    & 85.7 & 83.8    \\
\multicolumn{1}{|l|}{}                       & 5 QA pairs 2 overlap     &                         & 82.4 & 86.7    & 86.7 & 81.9    \\
\multicolumn{1}{|l|}{}                       & 10 QA pairs              &                         & 83.3 &\textbf{ 87.6}    & 81.4 & 81.4    \\
\multicolumn{1}{|l|}{}                       & 10 QA pairs 5 overlap    &                         & 80   & 87.6    & 82.3 & 82.9    \\
\multicolumn{1}{|l|}{}                       & 10 Utterances 4 overlap  &                         & 84.3 & 85.7    & \textbf{90}   & 84.7    \\ \hline \hline
\multicolumn{1}{|l|}{\multirow{5}{*}{\begin{turn}{90}Audio\end{turn}}} & Full Interview           & \multirow{3}{*}{SVM}    & 68.1 & 66.6    & 68.1 & 68.1    \\
\multicolumn{1}{|l|}{}                       & Answers Summed           &                         & 70.5 & 72.8    & 79.5 & 82.8    \\
\multicolumn{1}{|l|}{}                       & Answers Concatenated     &                         & 60.9 & 58.1    & 58.5 & 60      \\ \cline{2-7} 
\multicolumn{1}{|l|}{}                       & Answers Chunks           & \multirow{2}{*}{BiLSTM} & 74.3 & 56.2    & 74.3 & 58.6    \\
\multicolumn{1}{|l|}{}                       & 10 Utterances 4 overlap  &                         & \textbf{83.3} & 80.5    & \textbf{85.7} & 80.1    \\ \hline \hline
\multicolumn{1}{|l|}{\multirow{3}{*}{\begin{turn}{90}Video\end{turn}}} &
  Answers only All Features &
  \multirow{3}{*}{BiLSTM} &
  66.1 &
  68.1 &
  60 &
  68 \\
\multicolumn{1}{|l|}{}                       & Answers only 21 Features &                         & 51.4 & 63.8    & 54.7 & 66.6    \\
\multicolumn{1}{|l|}{}                       & Full Interview Features  &                         & \textbf{70.5} & 62.8    & 60   & \textbf{68.5}    \\ \hline
\end{tabular}
\end{table}

\subsubsection{Text Formats Evaluation}


Table \ref{tab:format-eval} summarizes the results of testing various text input formats for modeling depression and PTSD. For the whole interview format, both unstructured and QA-structured approaches were evaluated. In the chunking format, Question-Answer (QA) pairs are grouped differently: 5 QA pairs, 5 QA pairs 2 overlap, 10 QA pairs, and 10 QA pairs 5 overlap . Additionally, the 10 Utterances 4 overlap format groups utterances into sets of 10 with a 4-utterance overlap, independent of speaker identity.

For depression, the BiLSTM model with the 10 QA pairs 5 overlap format achieved the highest balanced accuracy (87.6 BA), followed closely by 10 Utterances 4 overlap (85.7 BA) and 10 QA pairs (87.6 BA). Data normalization generally improved results, emphasizing the importance of preserving conversational flow for detecting depression.
For PTSD, the 10 Utterances 4 overlap format with the BiLSTM model notably outperformed all others, achieving a balanced accuracy of (90.0 BA). 5 QA pairs 2 overlap (86.7 BA) and 5 QA pairs (85.7 BA) followed closely behind. Smaller chunk sizes with overlap were effective, but notably, the utterance-based format showed superior results. Normalization generally reduced performance for PTSD, suggesting it may not suit PTSD detection. These findings demonstrate that chunking strategies, particularly utterance-based groupings with overlap, significantly impact performance, varying by psychological condition.
\subsubsection{Audio Formats Evaluation}
In exploring audio data for mental health analysis, various input formats are tested to optimize model performance, as summarized in Table \ref{tab:format-eval}. The initial approach utilizes the full unstructured audio of the interview. Alternative methods involve manipulating the audio responses of the participants, with one format summing the embeddings at each timestep into a single vector (Answers Summed), and another concatenating all responses (Answers Concatenated) into one continuous audio file before embedding. Another approach segments each answer as an individual timestep (Answers Chunks) using BiLSTM. Additionally, the 10 Utterances 4 overlap format groups utterances into sets of 10 with a 4-utterance overlap, independent of speaker identity.

For depression detection, the BiLSTM model with the 10 Utterances 4 overlap format achieves the highest balanced accuracy (83.3\% BA without normalization and 80.5\% BA with normalization). Data normalization generally reduces accuracy across audio-based methods for depression, suggesting that normalization may disrupt meaningful acoustic patterns important for this condition.

For PTSD, similarly, the 10 Utterances 4 overlap format outperforms other formats, achieving the highest balanced accuracy (85.7\% BA without normalization and 80.1\% BA with normalization). The Answers Summed format using SVM follows closely (82.8\% BA normalized), while concatenation and chunking approaches perform comparatively worse. Again, normalization slightly decreases performance, implying that preserving original audio characteristics may be beneficial for PTSD detection. These results highlight that utterance-based segmentation with overlap significantly improves classification accuracy for audio-based detection of depression and PTSD.
\subsubsection{Video Formats Evaluation}
In the analysis of video data for mental health detection, three distinct feature sets are explored, as detailed in Table \ref{tab:format-eval}. The first format (Answers only All Features) utilizes all available OpenFace features extracted exclusively from participant answers. The second approach (Answers only 21 Features) is more selective, using only 21 specific features identified by Santosh V. Patapati \cite{patapati2024integratinglargelanguagemodels}, chosen based on their demonstrated relevance and effectiveness in prior research. The third format (Full Interview Features) includes the complete set of OpenFace features extracted from the answers within the context of the full interview.

For depression detection, extracting Full Interview Features yields the highest balanced accuracy without normalization (70.5\% BA). However, normalizing data notably improves performance for the selective Answers only 21 Features format (63.8\% BA normalized), despite it having initially lower accuracy. This indicates that targeted normalization might help enhance predictive signals for depression when feature selection is highly specific.
In the context of PTSD, the Full Interview Features approach achieves the highest balanced accuracy after normalization (68.5\% BA), closely followed by the Answers only All Features format (68.0\% BA normalized). The selective Answers only 21 Features format demonstrates lower performance (66.6\% BA normalized), suggesting that utilizing a broader range of OpenFace features captures more relevant behavioral signals for PTSD detection. Normalization consistently improves accuracy across all feature sets for PTSD, underscoring its value in refining the predictive capability of video-based features. These findings emphasize the importance of comprehensive feature extraction combined with normalization when detecting PTSD from video data.
\subsection{Embedding Model Evaluation}
This section details the evaluation of various embedding models, specifically for text and audio modalities. The purpose of this evaluation is to identify the most effective embedding model for each modality, which will then be utilized in subsequent experiments. As pre-extracted video features are provided with the dataset, the evaluation focuses solely on text and audio embedding models.

\begin{table}[ht]
\centering
\caption{Performance Comparison of Embedding Models for Depression and PTSD Prediction Using Balanced Accuracy. Results are ranked from best to worst, with values in bold indicating the highest balanced accuracy achieved by the top-performing model.}
\label{tab:combined-table}

\begin{minipage}{0.48\textwidth} 
  \centering
  \caption*{(a) Text Embedding Models}
  \label{tab:text-emb}
  \begin{tabular}{cccc}
    \toprule
    Model                                   & Vector Size & Depression & PTSD \\ \midrule
    Text-embeddings-small                   & 1536        & \textbf{79 }        & \textbf{80}   \\
    Text-embdeding-004                      & 768         & 78.5       & 77.6 \\
    Text-embdeding-005                      & 768         & 78.1       & 75.7 \\
    Text-embeddings-large                   & 3072        & 70.9       & 77.6 \\
    NV-Embed-V2 - 8 bit                     & 4096        & 71.4       & 71.4 \\
    NV-Embed-V2 - 4 bit                     & 4096        & 78.1       & 70.9 \\
    KaLM                                    & 896         & 67.1       & 70.9 \\
    Stella                                  & 1024        & 67.1       & 71   \\ \bottomrule
  \end{tabular}
\end{minipage}\hfill 
\begin{minipage}{0.48\textwidth} 
  \centering
  \caption*{(b) Audio Embedding Models (SVM)}
  \label{tab:audio_embeddings}
  \begin{tabular}{lcc}
    \toprule
    Approach         & Vector Size & BA        \\ \midrule
    Whisper-small    & 768         & \textbf{68.1}      \\
    Whisper-large-turbo & 1280        & 68.1      \\
    wav2vec          & 768         & 66.19     \\
    Whisper-large    & 1280        & 65.24     \\
    wav2vec-large    & 1024        & 62.86     \\
    Whisper-base     & 512         & 59.05     \\
     \bottomrule
  \end{tabular}
\end{minipage}

\end{table}

\subsubsection{Text Embedding Models Evaluation}
A comprehensive evaluation of both open-source and closed-source text embedding models was conducted. Closed-source models tested include OpenAI's `text-embedding-small` and `text-embedding-large`, and Google's `text-embeddings-004` and `text-embeddings-005`. For open-source models, Nvidia's `NV-embed-V2`, `KaLM`, and `stella` were evaluated. This diverse set of models allows for a comparison of model size, architecture, and performance across different sources.
Table \ref{tab:text-emb} (a) presents the comparative results of various text embedding models for the prediction of depression and PTSD. The models were evaluated based on their balanced accuracy (BA) scores, and the table also lists the vector size for each model.  

The `text-embedding-small` model achieved the highest BA for both depression depression with a score of 79\% and 80\% respectively , followed closely by `text-embedding-004` with 78.5\%  for Depression and 77.6 for PTSD. The `text-embeddings-005` and `NV-Embed-V2- 4 bit` models also showed good performance, both achieving 78.1\% in Depression.  In contrast, `KaLM` and `stella` performed significantly lower, with a BA of 67.1\% for Depression and 71 for PTSD.

Overall, these results suggest that while models like  `text-embedding-small` offered superior performance for both depression and PTSD, smaller models such as `text-embedding-004` and `text-embdeding-005`  with vector sizes of 768 also performed competitively for the two tasks. This highlights a trade-off between model complexity and performance in text embedding models. Furthermore, the significantly lower performance of `KaLM` and `stella` indicates that not all models are equally effective, emphasizing the importance of empirical evaluation when choosing embedding models for specific tasks.
\subsubsection{Audio Embedding Models Evaluation}


Table \ref{tab:audio_embeddings} (b) presents the performance of various audio embedding models for predicting Depression using the SVM classifier. Among the models, 'Whisper-small' and 'Whisper-large-turbo' achieved the highest accuracy of 68.1\%, demonstrating their effectiveness in extracting meaningful audio features for this classification task. In contrast, the 'Whisper-base' model, with a smaller vector size of 512, achieved the lowest accuracy of 59.05\%. The `wav2vec` model with a vector size of 768 performed reasonably well, achieving an accuracy of 66.19\%, whereas its larger variant, 'wav2vec-large', with a vector size of 1024, performed slightly worse at 62.86\%, suggesting that simply increasing the model size does not always guarantee improved performance. Overall, these results highlight that Whisper-based embeddings, particularly 'Whisper-small' and 'Whisper-large-turbo', are highly effective for audio-based predictions of depression and anxiety, making them strong candidates for further research and practical applications.

\subsection{Feature Extraction}
This section presents the feature extraction process across text, audio, and video modalities using CNN and CNN-BiLSTM models. The goal is to determine the most effective configurations for each modality, which are then applied to depression and PTSD classification. Various chunking strategies are evaluated to assess their ability to capture relevant temporal, contextual, and multimodal features. The following subsections outline the performance of different approaches and highlight the most effective strategies for each modality.
\begin{table}[ht]
\centering
\caption{Single Modality Evaluation for Severity and Multi-Class Classification Using CNN and CNN-BiLSTM Hybrid Architectures Across Text, Audio, and Video Modalities. Severity is evaluated using MAE, and multi-class classification is assessed using balanced accuracy. Values in bold represent the highest balanced accuracy achieved for each approach.}
\label{tab:feats-ex}
\begin{tabular}{|ll|l|cc|cc|}
\hline
\multicolumn{2}{|c|}{\multirow{2}{*}{Approach}} &
  \multicolumn{1}{c|}{\multirow{2}{*}{\textbf{Model}}} &
  \multicolumn{2}{c|}{\textbf{Depression}} &
  \multicolumn{2}{c|}{\textbf{PTSD}} \\ \cline{4-7} 
\multicolumn{2}{|c|}{} &
  \multicolumn{1}{c|}{} &
  \textbf{BA} &
  \textbf{BA (w/ norm)} &
  \textbf{BA} &
  \textbf{BA (w/ norm)} \\ \hline
\multicolumn{1}{|l|}{\multirow{4}{*}{\begin{turn}{90}Text\end{turn}}} &
  \multirow{2}{*}{5 QA pairs 2 overlap} &
  CNN-BILSTM &
  \textbf{89} &
  86.7 &
  83.8 &
  84.7 \\
\multicolumn{1}{|l|}{} &
   &
  CNN &
  86.2 &
  86.2 &
  87.6 &
  84.8 \\ \cline{2-7} 
\multicolumn{1}{|l|}{} &
  \multirow{2}{*}{10 Utterances 4 overlap} &
  CNN-BILSTM &
  87.1 &
  86.2 &
  \textbf{90.5} &
  88.6 \\
\multicolumn{1}{|l|}{} &
   &
  CNN &
  83.3 &
  83.3 &
  83.8 &
  87.1 \\ \hline \hline
\multicolumn{1}{|l|}{\multirow{4}{*}{\begin{turn}{90}Audio\end{turn}}} &
  \multirow{2}{*}{Answers Chunks} &
  CNN &
  71.4 &
  69 &
  73.3 &
  76.2 \\
\multicolumn{1}{|l|}{} &
   &
  CNN-BILSTM &
  60.5 &
  60 &
  74.3 &
  70.5 \\ \cline{2-7} 
\multicolumn{1}{|l|}{} &
  \multirow{2}{*}{10 Utterances 4 overlap} &
  CNN &
  76.6 &
  80 &
  71.9 &
  75.2 \\
\multicolumn{1}{|l|}{} &
   &
  CNN-BILSTM &
  73.3 &
  \textbf{85.7} &
  78.6 &
  \textbf{85.2} \\ \hline \hline
\multicolumn{1}{|l|}{\multirow{4}{*}{\begin{turn}{90}Video\end{turn}}} &
  \multirow{2}{*}{Answers only All Features} &
  CNN &
  69.5 &
  66.1 &
  71.4 &
  68 \\
\multicolumn{1}{|l|}{} &
   &
  CNN-BILSTM &
  66.1 &
 \textbf{ 71.4} &
  58 &
  71.4 \\ \cline{2-7} 
\multicolumn{1}{|l|}{} &
  \multirow{2}{*}{Full Interview Features} &
  CNN &
  52.8 &
  65.2 &
  70.5 &
  72.8 \\
\multicolumn{1}{|l|}{} &
   &
  CNN-BILSTM &
  65.2 &
  69 &
  \textbf{73.8} &
  70 \\ \hline
\end{tabular}
\end{table}
\subsubsection{Text  Modality}
Feature extraction from text data identifies emotional and behavioral patterns associated with depression and PTSD. The most effective configurations evaluated involve the CNN and CNN-BiLSTM models under two chunking strategies, 5 QA pairs 2 overlap and 10 Utterances 4 overlap, summarized in Table \ref{tab:feats-ex}. For depression classification, the CNN-BiLSTM model under the 5 QA pairs 2 overlap configuration achieves the highest balanced accuracy (BA) of 89\%, capturing essential temporal dependencies. For PTSD detection, the CNN-BiLSTM model under the 10 Utterances 4 overlap configuration achieves the highest balanced accuracy (90.5\% without normalization and 88.6\% with normalization), demonstrating strong performance in capturing temporal and contextual information. The CNN model also performs well, notably achieving 87.6\% BA under the 5 QA pairs 2 overlap configuration. Overall, CNN and CNN-BiLSTM models enhance classification accuracy significantly, with the CNN-BiLSTM model using the 10 Utterances 4 overlap chunking strategy emerging as the optimal configuration for PTSD detection, while the 5 QA pairs 2 overlap strategy is preferable for depression.

\subsubsection{Audio Modality}
Audio feature extraction is analyzed using two chunking strategies, Answers Chunks and 10 Utterances 4 overlap, with CNN and CNN-BiLSTM models, as shown in Table \ref{tab:feats-ex}. For depression detection, the CNN-BiLSTM model with the 10 Utterances 4 overlap configuration achieves the highest balanced accuracy (85.7\% with normalization), highlighting the effectiveness of combining CNN-based feature extraction with sequential modeling. The CNN model under the same configuration achieves 80\% BA with normalization, further supporting the benefits of this chunking strategy. In contrast, the CNN-BiLSTM model with Answers Chunks performs the worst for depression, obtaining only 60.5\% BA without normalization.
For PTSD detection, the CNN-BiLSTM model using 10 Utterances 4 overlap with normalization achieves the highest balanced accuracy (85.2\%), reinforcing the effectiveness of this approach in capturing relevant audio features. The CNN model under Answers Chunks achieves 76.2\% BA with normalization, showing moderate performance. Overall, CNN-BiLSTM with the 10 Utterances 4 overlap configuration consistently delivers superior results, particularly when normalization is applied, underscoring its significance in enhancing audio-based classification of depression and PTSD.


\subsubsection{Video Modality}
Video feature extraction is analyzed using two configurations, Answers only All Features and Full Interview Features, with CNN and CNN-BiLSTM models. For depression detection, the CNN model using (Answers only All Features) achieves the highest balanced accuracy (69.5\%), while the CNN-BiLSTM model improves with normalization, reaching 71.4\%. Overall, normalization enhances depression classification across the video modality, improving performance in most configurations.
For PTSD detection, the CNN-BiLSTM model using (Full Interview Features) achieves the highest balanced accuracy (73.8\%), demonstrating its effectiveness in capturing PTSD-related visual cues. Similarly, normalization consistently improves PTSD classification across different configurations, particularly benefiting CNN-BiLSTM with (Answers only All Features), which increases from 58\% to 71.4\%. These results suggest that normalization is beneficial for both depression and PTSD detection in video-based models, enhancing the predictive capacity of CNN-BiLSTM across different feature extraction strategies.


\subsection{Severity and Multi-Class}
\begin{table}[ht]
\centering
\caption{Single Modality Evaluation for Severity and Multi-Class Classification Using CNN and CNN-BiLSTM Hybrid Architectures Across Text, Audio, and Video Modalities. Severity is evaluated using MAE and multi-class classification is assessed using balanced accuracy. Values in bold represent the highest balanced accuracy achieved for each approach.}
\label{tab:sev and multi}
\resizebox{\textwidth}{!}{%
\begin{tabular}{|lccccccc|}
\hline
\multicolumn{1}{|l|}{\multirow{2}{*}{Format}} &
  \multicolumn{1}{c|}{\multirow{2}{*}{Eval Metric}} &
  \multicolumn{2}{c|}{Text} &
  \multicolumn{2}{c|}{Audio} &
  \multicolumn{2}{c|}{Video} \\ \cline{3-8} 
\multicolumn{1}{|l|}{} &
  \multicolumn{1}{c|}{} &
  CNN &
  \multicolumn{1}{c|}{CNN-BiLSTM} &
  CNN &
  \multicolumn{1}{c|}{CNN-BiLSTM} &
  CNN &
  CNN-BiLSTM \\ \hline
\multicolumn{8}{|c|}{Depression} \\ \hline
\multicolumn{1}{|l|}{10 Utterances 4 overlap} &
  \multicolumn{1}{c|}{\multirow{2}{*}{MAE}} &
  0.62 &
  \multicolumn{1}{c|}{\textbf{0.59}} &
  0.85 &
  \multicolumn{1}{c|}{0.82} &
  0.98 &
  0.92 \\
\multicolumn{1}{|l|}{10 Utterances 4 overlap w/ norm} &
  \multicolumn{1}{c|}{} &
  0.62 &
  \multicolumn{1}{c|}{0.63} &
  0.75 &
  \multicolumn{1}{c|}{\textbf{0.66}} &
  0.92 &
  \textbf{0.84} \\ \hline
\multicolumn{8}{|c|}{PTSD} \\ \hline
\multicolumn{1}{|l|}{10 Utterances 4 overlap} &
  \multicolumn{1}{c|}{\multirow{2}{*}{MAE}} &
  0.37 &
  \multicolumn{1}{c|}{0.35} &
  0.53 &
  \multicolumn{1}{c|}{0.62} &
  0.8 &
  0.59 \\
\multicolumn{1}{|l|}{10 Utterances 4 overlap w/ norm} &
  \multicolumn{1}{c|}{} &
  0.36 &
  \multicolumn{1}{c|}{\textbf{0.26}} &
  0.59 &
  \multicolumn{1}{c|}{\textbf{0.39}} &
  0.78 &
  \textbf{0.62} \\ \hline
\multicolumn{8}{|c|}{Multi-Class} \\ \hline
\multicolumn{1}{|l|}{10 Utterances 4 overlap} &
  \multicolumn{1}{c|}{\multirow{2}{*}{BA}} &
  63.5 &
  \multicolumn{1}{c|}{\textbf{67.5}} &
  51.1 &
  \multicolumn{1}{c|}{\textbf{58.8}} &
  45 &
  \textbf{56} \\
\multicolumn{1}{|l|}{10 Utterances 4 overlap w/ norm} &
  \multicolumn{1}{c|}{} &
  58.3 &
  \multicolumn{1}{c|}{62.2} &
  49.1 &
  \multicolumn{1}{c|}{49.7} &
  51.1 &
  53.3 \\ \hline
\end{tabular}
}
\end{table}
This section extends the analysis beyond the binary classification of depression and PTSD presence/absence to encompass more nuanced aspects of mental health assessment. Specifically, it explores two additional tasks: severity prediction and multi-class classification. Severity prediction aims to estimate the degree of depression or PTSD experienced by an individual, providing a more granular assessment than a simple binary label. This is crucial for understanding the intensity of symptoms and potentially tailoring interventions. Multi-class classification, on the other hand, moves beyond individual diagnoses to consider the possibility of co-occurring conditions. This addresses the reality that individuals may experience both depression and PTSD simultaneously, or neither condition. By examining these more complex scenarios, the analysis aims to provide a more comprehensive and clinically relevant evaluation of the models' capabilities.
\subsubsection{Severity}
Performance for severity prediction is evaluated using Mean Absolute Error (MAE), where lower values indicate better performance (a smaller difference between predicted and actual severity scores). The 10 Utterances 4 overlap format(the best preforming format in the binary classification experiments), both with and without normalization, is used for evaluation across text, audio, and video modalities, employing both CNN and CNN-BiLSTM models in Table \ref{tab:sev and multi}.

For depression severity prediction, the text modality demonstrates the best performance, achieving a MAE of 0.62 for the CNN and 0.59 for the CNN-BiLSTM. The Audio modality has higher MAEs of 0.85 with CNN and 0.82 with CNN-BiLSTM, indicating less accurate predictions. The Video modality exhibits the highest MAEs, 0.98 (CNN) and 0.92 (CNN-BiLSTM). Normalization slightly influences performance. In the text modality, normalization shows no significant change. For audio, normalization reduces MAE to 0.75 (CNN) and 0.66 (CNN-BiLSTM), suggesting a benefit from normalization. In video, normalization reduces the MAE for the CNN-BiLSTM model to 0.84.

For PTSD severity prediction, the text modality again achieves the lowest MAE (best performance), with values of 0.37 (CNN) and 0.35 (CNN-BiLSTM). Normalization further improves performance, reducing the MAEs to 0.36 and 0.26, respectively. The audio and video modalities exhibit higher MAEs, indicating less accurate severity predictions for these modalities.
\subsubsection{Multi-Class}
Multi-class classification performance is measured using Balanced Accuracy (BA). As shown in Table \ref{tab:sev and multi}, the text modality achieves the highest Balanced Accuracy (BA), with 63.5\% for the CNN and 67.5\% for the CNN-BiLSTM model. The audio modality follows with 51.1\% (CNN) and 58.8\% (CNN-BiLSTM). The video modality shows the lowest performance, with a BA of 45\% (CNN) and 56\% (CNN-BiLSTM). These results indicate that the text modality provides the most discriminative information for multi-class classification in this context, although the audio modality using CNN-BiLSTM also demonstrates competitive performance.
\subsection{Classification}

\begin{figure}[!ht]
    \centering
    \begin{subfigure}{0.45\textwidth}
        \centering
        \includegraphics[width=\linewidth]{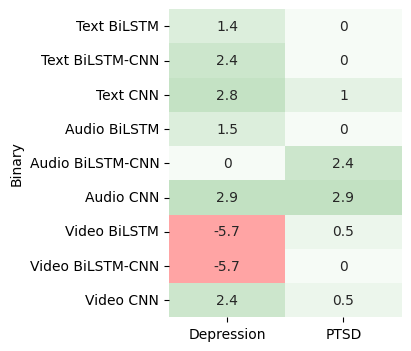}
        \label{fig:bin}
    \end{subfigure}
    \hfill
    \begin{subfigure}{0.3\textwidth}
        \centering
        \includegraphics[width=\linewidth]{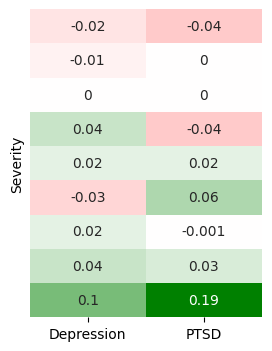}
        \label{fig:sev}
    \end{subfigure}
    \hfill
    \begin{subfigure}{0.24\textwidth}
        \centering
        \includegraphics[width=\linewidth]{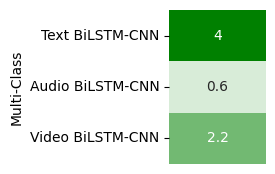}
        \label{fig:multi}
    \end{subfigure}
    \caption{Comparison using MLP Classification layer vs SVM Classifier. For Binary and Multi-class classification the metric used BA, the MAE for severity. The
heatmap illustrate the performance change when switching from an MLP to an SVM classifier.}
    \label{fig:all}
\end{figure}

This section investigates the impact of replacing the MLP classification layer with an SVM on model performance. As described in Section \ref{class}, the deep learning models (CNN and BiLSTM variations) are used for feature extraction, with the final MLP layer replaced by an SVM during inference. The following analysis examines the change in performance across binary classification, severity prediction, and multi-class classification tasks. The reported values represent the change between using SVM or MLP, for each modality and model. For Binary classification and Multi-class the metric used is the Balanced Accuracy, the MAE for severity.
The heatmap visualizations illustrate the performance change when switching from an MLP to an SVM classifier. Green cells indicate an improvement with the SVM (increased accuracy or decreased MAE), while red cells indicate a decrease in performance.

As shown in Figure \ref{fig:all},For depression, the impact of using an SVM varies across modalities. Text BiLSTM shows a small improvement (+1.4), while Text CNN experiences a slight increase (+2.8). Audio BiLSTM sees a modest improvement (+1.5), while Audio CNN has a bigger improvement (+2.9). Video BiLSTM and Video BiLSTM-CNN show notable decreases in performance (-5.7).
For PTSD, most changes are relatively small. Text models and audio BiLSTM remain nearly unchanged. Audio BiLSTM-CNN and Video BiLSTM-CNN see small improvements (+2.4 and 0, respectively). Video CNN has a little improvement (+0.5), while Audio CNN experiences the largest improvement(+2.9).

For depression severity, changes are generally small. Text and Audio BiLSTM models show slight improvements with the SVM, with increases close to zero. Video BiLSTM sees a small positive changes (+0.02)., while Video CNN and Audio BiLSTM-CNN show slight worsening (-0.03). Audio CNN show an improvement.
For PTSD severity, the results are mixed. Text BiLSTM and Text CNN show little change. Audio models shows small improvements, while Video models show mixed small changes.

The multi-class classification results, show the accuracy improvements using an SVM instead of MLP. Text BiLSTM-CNN demonstrates the biggest increase in accuracy (+4), followed by Video BiLSTM-CNN (+2.2). Audio BiLSTM-CNN shows a smaller improvement (+0.6).

Overall, the impact of switching to an SVM classifier is varied and depends on the specific task, modality, and model architecture. While some configurations show improvements in accuracy or reductions in MAE, others experience performance decreases. The most substantial improvements are seen in the multi-class cases and some improvements in Binary. These results suggest that SVMs can be a viable alternative to MLP classifiers in specific cases.

\subsection{Data Fusion}
This section evaluates different data fusion strategies for depression and PTSD classification, assessing their impact on model performance. Three fusion levels are examined: data-level fusion, where modalities are combined before feature extraction; feature-level fusion, where extracted feature embeddings are concatenated; and decision-level fusion, where predictions from separate modality-specific models are aggregated. The goal of this evaluation is to determine the most effective fusion approach for improving classification accuracy. The following subsections present the results for each fusion method, highlighting the most effective configurations and their contributions to overall performance.
\subsubsection{Data-level Fusion}
\begin{table}[ht]
\centering
\caption{Data-Level Fusion Evaluation Using Balanced Accuracy for Depression and PTSD Classification. Values in bold represent the highest balanced accuracy achieved for each model.}
\label{tab:data fusion}
\begin{tabular}{|l|l|rr|rr|}
\hline
\multirow{2}{*}{Model} & \multirow{2}{*}{Format} & \multicolumn{2}{l|}{Depression}                    & \multicolumn{2}{l|}{PTSD}                          \\ \cline{3-6} 
                       &                         & \multicolumn{1}{l}{MLP} & \multicolumn{1}{l|}{SVM} & \multicolumn{1}{l}{MLP} & \multicolumn{1}{l|}{SVM} \\ \hline
\multirow{2}{*}{CNN}        & 10 Utterances 4 overlap      & \textbf{83.3}          & 81.9 & 65.7          & 66.7 \\
                            & 10 Utterances 4 overlap w/ norm & 80            & 79.5 & \textbf{85.7} & 83.8 \\ \hline
\multirow{2}{*}{CNN-BiLSTM} & 10 Utterances 4 overlap      & 72.4          & 66.7 & 70            & 76.6 \\
                            & 10 Utterances 4 overlap w/ norm & \textbf{85.2} & 81.9 & 79.5          & \textbf{83.8} \\ \hline
\end{tabular}
\end{table}

Table \ref{tab:data fusion} presents the results of data-level fusion, where text, audio, and video data are combined prior to feature extraction, as described in Section \ref{fusion}.1. The fused data is then processed using either a CNN or CNN-BiLSTM model, and classification is performed using either an MLP or an SVM. The 10 Utterances 4 overlap format, with and without normalization, is used. Balanced Accuracy (BA) is the evaluation metric.

For depression detection, the CNN-BiLSTM model with normalization achieves the highest BA using an MLP (85.2\%). However, switching to an SVM classifier reduces performance in all cases. The CNN model achieves 83.3\% BA without normalization and 80\% BA with normalization using MLP. For PTSD detection, the CNN model with normalization achieves the best performance using MLP(85.7\% BA).
Switching to an SVM classifier improves performance in some cases, most notably with the CNN-BiLSTM model without normalization, increasing the BA from 70\% to 76.6\%.
These results indicate that the effectiveness of data-level fusion and the choice of classifier (MLP vs. SVM) are dependent on the specific condition (depression or PTSD) and the model architecture (CNN or CNN-BiLSTM). Normalization's impact also varies, improving performance in some scenarios and reducing it in others. Overall data level fusion shows improvements for the PTSD detection.

\subsubsection{Feature-Level Fusion}

Table \ref{tab:feature fusion} presents the results of feature-level fusion, where the feature embeddings extracted separately from each modality (text, audio, and video) are concatenated, as described in Section \ref{fusion}.2. Two training approaches are compared: full training of the model (without pretraining) and using a pretrained model with a frozen backbone. The 10 Utterances 4 overlap format is used for all modalities. Both MLP and SVM classifiers are evaluated, and performance is measured using Balanced Accuracy (BA).

With full training (no pretraining), the MLP classifier achieves a BA of 78.1\% for depression and 81.4\% for PTSD. Switching to an SVM classifier results in a slight increase for depression (78.6\%) and PTSD (81.9\%). While using a pretrained model with a frozen backbone significantly improves performance. The MLP classifier achieves 88.6\% BA for depression and 92.9\% BA for PTSD. The SVM classifier achieves the same performance for both conditions.

These results demonstrate that feature-level fusion can be effective, particularly when using a pretrained model. The choice between MLP and SVM classifiers has a negligible impact when using the pretrained model with a frozen backbone, achieving identical and high BAs for both conditions. The pretrained approach demonstrates strong capabilities for both depression and PTSD detection, suggesting the pretraining with the frozen backbone has a great impact.

\begin{table}[ht]
\centering
\caption{Feature-Level Fusion evaluation using balanced accuracy for Depression and PTSD classification.Values in bold represent the highest balanced accuracy achieved}
\label{tab:feature fusion}
\begin{tabular}{|l|l|lr|lr|}
\hline
\multirow{2}{*}{Model}         & \multirow{2}{*}{Data Format} & \multicolumn{2}{l|}{Depression}          & \multicolumn{2}{l|}{PTSD}                \\ \cline{3-6} 
 &  & MLP & \multicolumn{1}{l|}{SVM} & MLP & \multicolumn{1}{l|}{SVM} \\ \hline
Full Training (No Pretraining) & 10 Utterances 4 overlap                   & \multicolumn{1}{r}{78.1} & 78.6 & \multicolumn{1}{r}{81.4} & 81.9 \\ \hline
Pretrained (Frozen Backbone)   & 10 Utterances 4 overlap                   & \multicolumn{1}{r}{88.6} & \textbf{88.6} & \multicolumn{1}{r}{92.9} & \textbf{92.9} \\ \hline
\end{tabular}
\end{table}

\subsubsection{Decision-Level Fusion}

Table \ref{tab:decision fusion} presents the results of decision-level fusion, where the binary predictions from separate modality-specific models are combined, as described in Section \ref{fusion}.3. Initial experiments with logit fusion yielded low performance; therefore, only the results for binary prediction fusion are reported. Two combinations of best performing modality-specific formats are evaluated:

\begin{itemize}
    \item \textbf{Fusion Format 1:} Text (10 Utterances 4 overlap), Audio (10 Utterances 4 overlap w/ norm), Video (10 Utterances 4 overlap w/ norm)
    \item \textbf{Fusion Format 2:} Text (5 QA pairs 2 overlap), Audio (10 Utterances 4 overlap w/ norm), Video (10 Utterances 4 overlap w/ norm)
\end{itemize}

\begin{table}[h]
\centering
\caption{Decision-Level fusion evaluation using balanced accuracy for Depression and PTSD Classification.Values in bold represent the highest balanced accuracy achieved. }
\label{tab:decision fusion}
\begin{tabular}{|l|c|c|}
\hline
  &Depression    & PTSD          \\ \hline
Fusion Format 1       & 90.5          & \textbf{94.3} \\
Fusion Format 1 + LLM & \textbf{91}   & 94.3          \\ \hline
Fusion Format 2       & 92.4          & \textbf{96.2}          \\
Fusion Format 2 + LLM & \textbf{94.8} & 96.2          \\ \hline
\end{tabular}
\end{table}

The impact of incorporating predictions from an LLM, specifically DeepSeek LLM predictions from \cite{ali2024leveragingaudiotextmodalities}, is also assessed ("+ LLM" rows). The binary predictions from each modality and, optionally, the LLM are concatenated and used as input to an SVM classifier. Balanced Accuracy (BA) is used as the evaluation metric.

For Fusion Format 1, the BA is 90.5\% for depression and 94.3\% for PTSD. Adding LLM predictions slightly improves depression detection (91\%) but maintains the same performance for PTSD (94.3\%).
For Fusion Format 2, a higher baseline performance is observed, with a BA of 92.4\% for depression and 96.2\% for PTSD. Adding LLM predictions further improves depression detection to 94.8\%, while PTSD detection performance remains unchanged.

These results demonstrate that decision-level fusion, particularly using binary prediction fusion, can achieve high accuracy for both depression and PTSD detection. The Fusion Format 2 configuration, utilizing the 5 QA pairs 2 overlap format for text, consistently outperforms Fusion Format 1. The inclusion of LLM predictions provides a modest but consistent improvement for depression detection, highlighting the potential of leveraging external knowledge sources. The optimal decision level fusion reaches the highest balanced accuracy among all the different levels of fusions.

\subsubsection{Severity and Multi-class Fusion}

This section analyzes the impact of decision-level and feature-level fusion strategies on severity prediction and multi-class classification. The results show that feature-level fusion performs better, as it captures richer contextual information, leading to improved accuracy. The findings highlight the importance of choosing the right fusion approach based on task complexity, particularly for distinguishing between multiple conditions.

Table \ref{tab:Sev and Multi fusion} presents the results of applying decision-level and feature-level fusion strategies to severity prediction and multi-class classification. This analysis aims to determine the most effective fusion approach for these more complex tasks, extending beyond the binary classification of depression and PTSD.

\begin{table}[ht]
\centering
\caption{Decision-Level and Feature-Level Fusion evaluation for Severity (using MAE) and Multi-class (using BA) problems.Values in bold represent the highest balanced accuracy achieved.}
\label{tab:Sev and Multi fusion}
\begin{tabular}{|l|l|l|cc|}
\hline
\multicolumn{1}{|c|}{Category} & \multicolumn{1}{c|}{Fusion Type}       & Mertic               & \multicolumn{1}{c|}{Depression}    & PTSD          \\ \hline
\multirow{3}{*}{Severity}      & Decision-Level - Fusion Format 1       & \multirow{3}{*}{MAE} & \multicolumn{1}{c|}{0.55}          & 0.25          \\
                               & Decision-Level - Fusion Format 1 + LLM &                      & \multicolumn{1}{c|}{\textbf{0.48}} & \textbf{0.23} \\
 & Feature-Level - Pretrained (Frozen Backbone) &  & \multicolumn{1}{c|}{0.50} & 0.27 \\ \hline
\multirow{3}{*}{Multi-Class}   & Decision-Level - Fusion Format 1       & \multirow{3}{*}{BA}  & \multicolumn{2}{c|}{74}                            \\
 & Decision-Level - Fusion Format 1 + LLM       &  & \multicolumn{2}{c|}{\textbf{78}} \\
 & Feature-Level - Pretrained (Frozen Backbone) &  & \multicolumn{2}{c|}{77.5}        \\ \hline
\end{tabular}
\end{table}

For severity prediction, measured by Mean Absolute Error (MAE), decision-level fusion generally outperforms feature-level fusion. The incorporation of LLM predictions (DeepSeek LLM from \cite{ali2024leveragingaudiotextmodalities}) within the decision-level fusion framework (Fusion Format 1: text Utter 10-4, audio Utter 10-4 Norm, video Utter 10-4 Norm) consistently improves performance, lowering the MAE for both depression (from 0.55 to 0.48) and PTSD (from 0.25 to 0.23). This suggests that the LLM provides additional, relevant information that helps refine the severity estimates. Feature-level fusion using a pretrained model with a frozen backbone shows slightly worse results than decision level fusion, achieving an MAE of 0.5 for depression and 0.27 for PTSD.

For multi-class classification, decision-level fusion (Fusion Format 1) shows the highest balanced accuracy (78\%), exceeding the feature-level fusion approach and the decision level fusion without including the LLM (77.5\% and 74\% respectively). Similar to the observation in the previous section using Binary Prediction Fusion, combining modality predictions offers a significant advantage.

Overall, these findings demonstrate that decision-level fusion, particularly when augmented with LLM predictions, is a highly effective strategy for both severity prediction and multi-class classification. The ability to integrate information at the decision level, incorporating external knowledge from the LLM, allows the model to make more nuanced and accurate assessments of mental health conditions, going beyond simple binary classifications. The consistent improvement provided by the LLM across both tasks highlights its value in enhancing the performance of multimodal fusion systems for mental health analysis.

\subsection{ Comparative Evaluation}

Our proposed multimodal approach was evaluated against existing literature using the E-DAIC dataset for binary classification tasks. As shown in Table \ref{tab:comparsions}, our system achieves state-of-the-art performance, yielding an accuracy of 94.6\% for depression detection and a balanced accuracy of 96.2\% for PTSD detection. These results surpass all other reported metrics in the compared studies for these specific binary classification tasks, demonstrating the effectiveness of our method in accurately distinguishing between the presence and absence of these conditions based on the defined criteria (Accuracy (ACC) for depression, BA for PTSD).

Beyond binary classification, we extended our analysis to severity prediction and multi-class classification. To the best of our knowledge, directly comparable results for these tasks using the same severity thresholds and multi-class definitions on the E-DAIC test set are scarce in the literature, with the notable exception of the work by Abdelrahman A. Ali et al. \cite{ali2024leveragingaudiotextmodalities}. For a rigorous comparison, we selected their best-performing model configuration (considering both ZS and FS results presented) for each respective task (depression severity, PTSD severity, and multi-class) and evaluated its performance on the E-DAIC test set. As indicated in Table \ref{tab:comparsions}, our approach achieves considerably lower MAE values for both depression (0.48) and PTSD (0.23) severity prediction and a significantly higher balanced accuracy (78\%) for multi-class classification compared to their reported bests.

These results comprehensively highlight the effectiveness of our combination of utterance-level chunking, optimized embedding models, CNN-BiLSTM feature extraction, and, crucially, decision-level fusion incorporating LLM predictions. The substantial improvements across all evaluated tasks underscore the advantages of our holistic multimodal approach and its potential for developing more accurate, nuanced, and clinically relevant mental health assessment tools. 
\begin{table}[ht]
\centering
\caption{Comparative performance of our proposed method against existing studies for binary classification of depression and PTSD. Accuracy (ACC) is used for depression detection, while Balanced Accuracy (BA) is used for PTSD detection. '-' denotes that the result was not reported in the original publication, or the study did not evaluate that condition. Values in bold represent the highest balanced accuracy achieved.}
\label{tab:comparsions}
\resizebox{\textwidth}{!}{%
\begin{tabular}{cccccc}
\hline
 Method &
  Depression (ACC) &
  PTSD (BA) &
  Depression Severity (MAE) &
  PTSD Severity (MAE) &
  Multiclass (BA) \\ \hline
Ours                      & \textbf{94.6\%}  & \textbf{96.2\%}& \textbf{0.48} & \textbf{0.23} & \textbf{78\%}  \\
Farah Mohammad et al. \cite{MOHAMMAD20244125}     & 91.74\% & -               & -    & -    & -      \\
Mariia Nykoniuk et al. \cite{computation13010009}   & 86\%   & -               & -    & -    & -     \\
Xu Zhang et al. \cite{Zhang2025}          & 82\%    & -               & -    & -    & -       \\
Anup Kumar Gupta et al. \cite{GUPTA2024109325}   & 80.3\%  & -               & -    & -    & -      \\
Joseph Cameron et al. \cite{cameron2024multimodalgenderfairnessdepression}    & 70\%    & -               & -    & -    & -      \\
Wu et al. (CALLM)  \cite{10614318}       & -       & 77\% & -               & -    & -       \\
Abdelrahman A. Ali et al. \cite{ali2024leveragingaudiotextmodalities} with ZS & 75\%       & 76\%   & 0.71 & 0.48 &33\% \\ 
Abdelrahman A. Ali et al. \cite{ali2024leveragingaudiotextmodalities} with FS & 82.3\%       & 82.4\%   &  0.60& 0.42 &34.4\% \\
\hline
\end{tabular}%
}
\end{table}

\section{Limitations and Future Work}
While this study demonstrates the significant potential of multimodal machine learning for mental health assessment, several avenues for future research remain.  These can be broadly categorized into improvements to the current methodology, expansion of the scope, and exploration of real-world applications.

\textbf{Enhancing Model Architecture and Fusion Strategies:}
Multimodal machine learning is revolutionizing mental health assessment by integrating diverse data sources like text, audio, and physiological signals for a holistic view of an individual’s well-being. Advances in model architectures and techniques are paving the way for more accurate and clinically relevant solutions. Here, we highlight three promising directions transformer-based fusion, Explainable AI, and adaptive fusion to enhance performance and trust in these systems.
\begin{itemize}
    \item \textbf{Transformer-Based Fusion:}  
    While we explored CNN and BiLSTM architectures, the current state-of-the-art in many multimodal tasks involves Transformer models.  Investigating Transformer-based fusion mechanisms, such as cross-modal attention, could further improve the integration of information from different modalities and capture more complex relationships.
    \item \textbf{Explainable AI (XAI):}  Developing methods to explain the model's predictions is crucial for building trust and gaining clinical acceptance.  Future work should incorporate XAI techniques, such as attention visualization, feature importance analysis, and rule extraction, to provide insights into \textit{why} a particular prediction was made.  This is particularly important for high-stakes decisions in mental health.
    \item \textbf{Adaptive Fusion:}  The optimal fusion strategy may vary depending on the individual and the specific context.  Exploring adaptive fusion techniques, where the model dynamically learns how to weight and combine information from different modalities based on the input data, could lead to more personalized and accurate assessments.
\end{itemize}

\textbf{Refining LLM Integration:}

    We observed that LLMs can contribute to improved performance, particularly in decision-level fusion.  Further research should focus on optimizing prompt engineering strategies and exploring fine-tuning of LLMs specifically for mental health-related tasks. This can improve the quality of their text-based predictions and their integration with other modalities.

\textbf{Clinical Validation:}

The ultimate test of any diagnostic tool is its performance in a real-world clinical setting.  Future work should involve prospective studies, where the model is evaluated on new, unseen data collected in a clinical environment. This will provide a more realistic assessment of its accuracy, reliability, and clinical utility.

By addressing these future research directions, we can move closer to realizing the full potential of multimodal machine learning for improving mental health assessment, diagnosis, and treatment, ultimately leading to better outcomes for individuals struggling with mental illness.


\end{document}